\documentclass[showpacs, pra,twocolumn,preprintnumbers ,amsmath, amssymb, superscriptaddress, aps]{revtex4-2}
\usepackage{color}
\usepackage{amsmath,amssymb}
\usepackage{pifont}
\usepackage{amssymb}  
\usepackage{bbold}
\usepackage{float}
\usepackage{subfloat}
\usepackage[caption=false]{subfig}
\usepackage{tikz}
\usepackage{makecell}
\usepackage{subfig}
\usepackage{pifont}   
\usepackage{graphicx} 
\graphicspath{{Figures/}}
\usepackage{dcolumn}  
\usepackage{bm}       
\usepackage{multirow} 
\usepackage{placeins}
\usepackage[colorlinks]{hyperref}
\usepackage{mathtools}
\usepackage{appendix}

\def \be{\begin{align}}
	\def \ee{\end{align}}
\def \bea{\begin{eqnarray}}
	\def \eea{\end{eqnarray}}

\begin{document}
	\renewcommand{\thesection}{\arabic{section}}
	\renewcommand{\thesubsection}{\arabic{section}.\arabic{subsection}}
	\renewcommand{\thefigure}{\arabic{figure}}

	\title{Spin--valley--resolved tunneling through magnetic barriers in WSe$_2$}

	\author{Rachid El Aitouni}
	\affiliation{Laboratory of Theoretical Physics, Faculty of Sciences, Choua\"ib Doukkali University, PO Box 20, 24000 El Jadida, Morocco}
	
	\author{Clarence Cortes}
	\affiliation{Vicerrector\'ia de Investigaci\'on y Postgrado, Universidad de La Serena, La Serena 1700000, Chile}  
	\author{David Laroze}
	\affiliation{Instituto de Alta Investigación, Universidad de Tarapacá, Casilla 7D, Arica, Chile}
	\author{Ahmed Jellal}
	\email{a.jellal@ucd.ac.ma}
	\affiliation{Laboratory of Theoretical Physics, Faculty of Sciences, Choua\"ib Doukkali University, PO Box 20, 24000 El Jadida, Morocco}
	
	\begin{abstract}
		We investigate the influence of a magnetic field on the electronic properties of WS$e_2$ with a focus on spin-orbit coupling, spin and valley polarization, and conductance. We {solve} the eigenvalue equation analytically and {use} the continuity equation to determine the transmission probability based on current densities. {We calculate the} conductance using Büttiker formula. Our numerical results indicate that transmission through the $K$ valley is more likely than through the $K'$ valley. For both valleys, the Klein tunneling effect is clearly observed. The conductance is affected by an increase in the magnetic field {because it alters} the energy levels of fermions via the Zeeman effect. These modifications enable the confinement of fermions within the barrier. Spin and valley polarization are also influenced by the magnetic field. As the field intensity increases, it steers the fermions and determines which channel can cross the barrier. This adds another tool of controlling fermions, paving the way for relevant applications in valleytronics and valley filtering for information storage.

	\end{abstract}
	
	\pacs{72.80.Vp, 73.23.-b, 78.67.-n
		\\{\sc Key Words:}
		WSe$_2$ layer, magentic field, transmission, conductance, plarization, Zeeman effect.}
	\maketitle
	
	\section{Introduction}	\label{Intro}
	
	Recently, the development of new materials {with reduced dimensions for smaller electronic components} has attracted the attention of the scientific communities. Isolating the structure of graphene {in  2004} opened a revolutionary gate to a new generation of materials that came to be known as {two-dimensional (2D)} materials \cite{Novos2004}. Generally speaking, any material that measures less than one nanometer in thickness may be termed {a 2D material} \cite{nano}.
	Graphene, in particular, has attracted enormous attention due to its exceptional electronic properties, notably its extremely high electron mobility \cite{mobil1,mobil2}. However, the absence of a band gap between its valence and conduction bands \cite{prop} makes charge carriers difficult to control, as they can freely transition between bands \cite{zero,zero1}. To overcome this limitation, several confinement strategies have been proposed. Simple electrostatic barriers allow partial confinement \cite{rect}, but electrons incident normally can cross the barrier without reflection, a phenomenon known as Klein tunneling \cite{klien1,klien2,klienexp}. Time-dependent or oscillating barriers have also been explored, where photon exchange between the barrier and fermions modifies their energy and can trap particles in degenerate states \cite{oscil1,oscil2,Laserquasi,doublelaser,doubletemps,timepot,timepot2}, giving rise to the dynamic Stark effect \cite{Stark}. Magnetic barriers provide another route to control transport \cite{mag1,mag3,mag4,magneticfield,doublemag}, as the applied magnetic field generates Landau levels \cite{Landau} that restrict fermions to well-defined energy states. Moreover, hybrid configurations combining static, oscillating, and magnetic barriers further enrich the physics by enabling the formation of quasi-bound states \cite{doublelaser+magn,laser+mag}, whose coupling with bound states can lead to Fano resonances \cite{Fano,bis2}.

	Although graphene exhibits a remarkably high carrier mobility, making it {an excellent medium for charge transport}, it inherently faces restrictions in the context of valleytronics. This {limitation} stems from the fact that it has little to no inherent spin-orbit interaction strength \cite{Zerocouplage}, which {makes it inefficient for controlling} the spin and valley channels. 
	Consequently, {significant research efforts have been devoted to identifying other 2D materials that can overcome this limitation]
	}
	Some of the most important candidates identified in this pursuit are transition metal dichalcogenides (TMDCs), specifically molybdenum disulfide (MoS$_2$) \cite{MoS2,MoS22} and tungsten diselenide (WS$e_2$) \cite{WSe2,WS2}. Unlike graphene, both materials have a direct band-gap structure  
	with strong intrinsic spin-orbit interaction \cite{couplage1,couplage2}, resulting in a substantial overlap between the valley and spin channels.
	This, in turn, makes both spin- and valley-polarized transport channels {easier to control.} Although {these materials exhibit relatively lower electron mobility} \cite{mobMoS2,mobWSe2} compared to  graphene, they still hold 
	{significant potential for advancing spin- and valley-based electronics
		owing to their strong spin–orbit interaction, which opens new avenues for manipulating electron transport.}

	Motivated by the growing interest in the electronic properties of two-dimensional materials, we investigate quantum tunneling in monolayer WSe$_2$ in the presence of a magnetic barrier. The barrier is modeled by two ferromagnetic strips deposited on the WSe$_2$ sheet, which partition the system into three regions, with a uniform magnetic field applied only in the central region. This configuration allows us to examine how magnetic confinement affects charge transport across the structure.
	The low-energy electronic states of WSe$_2$ are described by two coupled components associated with the valence and conduction bands, each influenced by intrinsic spin–orbit coupling. The applied magnetic field interacts with both the spin and valley degrees of freedom, giving rise to field-induced energy shifts through Zeeman-like terms \cite{Zemman}, which significantly modify the fermionic dynamics. The corresponding wave functions are obtained analytically by solving the eigenvalue equation in each region.
	Transmission and reflection coefficients are determined by enforcing spinor  continuity at the interfaces between regions, while edge effects associated with the ferromagnetic strips are neglected by assuming them to be infinitely extended. From these solutions, the current density is evaluated, enabling the evaluation of the total transmission and reflection probabilities. Our results show that increasing the magnetic field strength suppresses transmission by enhancing the coupling between charge carriers and the field, {leading} to a reduction in conductance. Moreover, transport is found to be dominated by carriers from a single valley, with spin-resolved contributions exhibiting an alternating behavior. Despite this spin selectivity, the overall populations of spin-up and spin-down carriers remain nearly balanced, highlighting the interplay between magnetic confinement and spin–valley coupling in WSe$_2$.

	The paper is structured as follows. In Sec.~\ref{Theory}, we present the theoretical model along with the derivation of the eigenspinors associated with the three regions composing the system. In Sec.~\ref{TTCC}, we use the continuity of the eigenspinors at the interfaces to determine the transmission and reflection amplitudes. Subsequently, we employ the current density to explicitly determine the  transmission probabilities. From these, we compute the corresponding conductance
	and polarization. 
	In Sec.~\ref{Num}, we discuss our numerical results and provide a detailed interpretation. Finally, we conclude by summarizing our findings.

	\section{Theoritical Model}	\label{Theory}
	
\begin{figure}[ht!]
		\centering
		\includegraphics[scale=0.55]{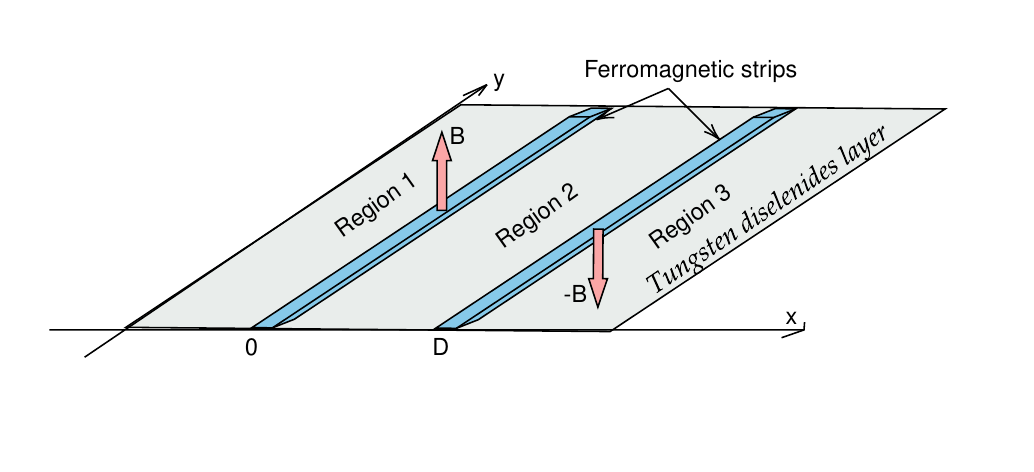}
		\caption{A schematic illustration of two ferromagnetic strips separated by a distance 
			$D$ and deposited on a WSe$_2$
			monolayer, defining three distinct regions due to the presence of the magnetic  barrier.
		}\label{str}
	\end{figure}

	Figure~\ref{str} illustrates the physical configuration of the present system, in which a magnetic barrier is formed within a monolayer of WSe$_2$. The barrier is created by depositing a ferromagnetic strip on top of the WSe$_2$ sheet, producing a localized magnetic field in the region under the strip while the surrounding areas remain field-free. This spatially inhomogeneous magnetic field divides the system into distinct regions and acts as an effective barrier for charge carriers. This setup allows the magnetic field to selectively influence the electronic states inside the barrier region, providing a controllable platform for studying spin- and valley-dependent transport properties in WSe$_2$.
	
	{In our model, the ferromagnetic barrier is treated as having a well-defined length along the transport direction, while its width in the transverse direction is considered much larger than the electronic wavelength. This allows us to neglect edge effects and treat the barrier as effectively one-dimensional. Such an approximation is valid when the length of the ferromagnetic strip is much greater than the electronic wavelength and the characteristic barrier width, ensuring that the main contribution to the transport comes from longitudinal scattering. Therefore, the qualitative features of the transmission and valley polarization discussed in this work are not significantly affected by the finite width of the ferromagnetic bar.}

	In the low-energy model, the Hamiltonian describing an electron in  WSe$_2$ is given by
	\begin{widetext}
			\begin{align}
				H =  v_F \left( \tau \sigma_x p_x + \sigma_y (p_y+eA_B \right) + \frac{\Delta}{2}\sigma_z + \lambda_c \tau s_z\frac{(\sigma_0+\sigma_z)}{2} +\lambda_v \tau s_z\frac{(\sigma_0-\sigma_z)}{2} +(s_zM_s-\tau M_v)\sigma_0
		\end{align}
	\end{widetext}
	where $(p_x,p_y)$ are the components of momentum vector, $\sigma_i$($i=x,y,z$) are the Pauli matrices, $\sigma_0$ is the unit matrix, $\tau=1(-1)$ denotes the ${K}({K'})$ valley,  $s_z=1(-1)$ represents spin up (down), $v_F=5\times10^5$m/s is the Fermi velocity, {$\Delta=1.7$eV is the band-gap, 
		$\lambda_{c,v}$ denote the spin-orbit coupling in both the conduction and valence bands, respectively, with  $\lambda_v=112.5$ meV and $\lambda_c=7.5$~meV~\cite{WSe2valeurs}. 
		The two parameters
		$M_s$ and $M_v$ are the perturbations created by the magnetic field, 
		known as the Zeeman effect \cite{Zemman}. They are defined as $M_{s,v} = \frac{g_{s,v} \, \mu_B \, B}{2}$, with $\mu_B = 5.788 \times 10^{-5} \, \text{eV/T}$ is the Bohr magneton. For WSe$_2$, the values of the spin and valley 
		$g$-factors are $g_s=2$ and $g_v=4$
		\cite{gs,gv}.
		The vector potential component
		$A_B$  associated with the magnetic field, created by two ferromagnetic strips,  takes the form  
		\begin{equation}
			A_B=Bl_B\left[\Theta(x)-\Theta(x-D)	\right]
		\end{equation}
		with  $l_B=\sqrt{\frac{\hbar}{eB}}$ is the magnetic length
		and $\Theta(x)$ denotes the step function.}

	The eigenspinors and eigenvalues can be obtained by solving $H \Psi = E \Psi$, where the spinor is decomposed as $\Psi(x,y,t)=(\psi_c(x),\psi_v(x))^\dagger e^{ik_yy}e^{-iEt/\hbar}$, which  includes the wave functions for the conduction (c) and valence (v) bands. To proceed further, let us  write the Hamiltonian in matrix form
	as
	\begin{align}\label{H}
		H =
		\begin{pmatrix}
			\Delta_c& v_F [\tau p_x - i (p_y+eA_B)] \\
			v_F [\tau p_x + i (p_y+eA_B)] & \Delta_v 
		\end{pmatrix}
	\end{align}
	where we have set 
	\begin{align}
		&\Delta_{c,v}=  \pm \frac{\Delta}{2}+\lambda_{c,v} \tau s_z+s_zM_s-\tau M_v\\
		& k_y^B=k_y +\frac{1}{l_B}. 
	\end{align}
	From the Hamiltonian  \eqref{H}, we show that 
	the eigenvalues are 
	\small\begin{align}\label{energie}
		&E^\mu_{s_z,\tau} =s_z M_s -  \tau M_v + \frac{1}{2} \left[s_z \tau (\lambda_c + \lambda_v) + 
		\mu\varepsilon_{s_z,\tau}\right]
	\end{align}
	where  $\varepsilon_{s_z,\tau}$ is given by
	\begin{equation}\label{eq7}
		\varepsilon_{s_z,\tau}=\sqrt{
			4 (\hbar v_F)^2 \left( k_x^2 + \left( k^B_y\right)^2 \right) + 
			\left( 2\Delta + s_z \tau (\lambda_c - \lambda_v) \right)^2}. 
	\end{equation}
	To determine the  corresponding eigenspinors, we use $H \Psi = E \Psi$
	to end up with two coupled equations
	\begin{align}
			&	v_F\left[i \tau \hbar\frac{\partial}{\partial x} - i \hbar k_y^B \right]\phi_v(x) = (E-\Delta_c)  \phi_c(x) \\
			&	v_F\left[i \tau \hbar\frac{\partial}{\partial x} + i \hbar k_y^B  \right]\phi_c(x) =(E-\Delta_v) \phi_v(x).
	\end{align}
	After solving the eigenvalue equations for each region (1,2,3), we obtain the following eigenspinors
	\begin{align}
			&\psi^1_{c,v}(x)=\left[\begin{pmatrix}
				1\\ e^{i\phi}
			\end{pmatrix}e^{i k_xx}+r\begin{pmatrix}
				1\\ -e^{-i\phi}
			\end{pmatrix}e^{-i k_xx}\right]
			\\
			&\psi^2_{c,v}(x)=\left[a_2\begin{pmatrix}
				1\\ e^{i\theta}
			\end{pmatrix}e^{i q_xx}+b_2\begin{pmatrix}
				1\\ -e^{-i\theta}
			\end{pmatrix}e^{-i q_xx}\right]
			\\
			&\psi^3_{c,v}(x)=\left[t\begin{pmatrix}
				1\\ e^{i\phi}
			\end{pmatrix}e^{i k_xx}\right]
	\end{align}
	where  the corresponding  wave vectors are 
	\begin{align}
			&k_x =\tau \sqrt{(E-\Delta_c) (E-\Delta_v)/(\hbar v_F)^2-k_y^2}\label{13}\\
			&q_x =\tau \sqrt{(E-\Delta_c) (E-\Delta_v)/(\hbar v_F)^2-(k_y^B)^2}\label{14} 
		\end{align}
	while the incident  angle is $\phi=\tan^{-1}\left(\frac{k_y}{k_x}\right)$ and emergent  angle  is $\theta=\tan^{-1}\left(\frac{k_y^B}{q_x}\right)$.
	{Note that the sign of the wave vector in~(\ref{13}-\ref{14})  is determined by the valley index. Specifically, we take $\tau=+1$ for carriers in the $K$ valley and $\tau=-1$ for carriers in the $K'$ valley. This choice ensures that the direction of propagation is consistently defined with respect to the valley degree of freedom and avoids any ambiguity in the calculation of transmission, reflection, and valley-dependent transport properties. All subsequent expressions and derivations are based on this sign convention.
	}
	
	The energy spectrum obtained above serves as the foundation for analyzing the transport properties of the system. In particular, it is used to determine the transmission probability by identifying the propagating and evanescent modes present in each region. From this spectrum, the dependence of transmission on the system parameters can be systematically explored, allowing us to investigate related transport quantities such as the conductance, and polarization effects.

	\section{Transmission and conductance}\label{TTCC}
	
	{Before proceeding, let us recall the following. Since the barrier width is negligible compared to the length of the ferromagnetic strips, the barrier can be regarded as infinite along the \(y\)-direction. This approximation eliminates edge effects and lateral boundary interference when analyzing the transport properties \cite{infini1,infini2,infini3}. Consequently, the continuity of the eigenspinor at the interfaces requires that }
	\begin{align}
		& \Psi_1(0,y,t)=\Psi_2(0,y,t) \\
		& \Psi_2(D,y,t)=\Psi_3(D,y,t).
	\end{align}
	Since each eigenspinor consists of two components, each interface provides two scalar continuity relations. Therefore, these four equations fully determine the relations between the incident, reflected, and transmitted amplitudes, forming the basis for calculating the transmission and reflection probabilities. They are given by
	\begin{align}
			&1+r=a+b\\
			&e^{i\phi}-re^{-i\phi}=ae^{i\theta}-be^{-i\theta}\\
			&te^{i k_xD}=ae^{i q_xD}+be^{-i q_xD}\\
			&t e^{i\phi} e^{i k_xD}=ae^{i\theta} e^{i q_x D}-be^{-i\theta} e^{-i q_x D}
		\end{align}
	The resulting system of coupled equations is {straightforward to solve analytically. Upon completing the calculations, we obtain} explicit expressions for the transmission and reflection coefficients, providing direct insight into the tunneling behavior of carriers across the barrier. 
	{For the case $q_x^2>0$, these coefficients take the forms }
	\begin{align}
			&t=\frac{e^{-i D k_x }
				\cos\phi\cos\theta }{
				\cos\phi\cos\theta\cos(D q_x ) - i \sin(D q_x ) (1-\sin\phi\sin\theta)
			}\label{21}\\
			&r=\frac{e^{i\phi}\sin(Dq_x )(\sin\phi-\sin\theta)}{
				\cos\phi\cos\theta\cos(D q_x ) - i \sin(D q_x ) (1-\sin\phi\sin\theta)}.\label{22}
		\end{align}
		When $q_x^2<0$, the longitudinal wave vector becomes imaginary and can be written as $q_x=i\kappa$, where $\kappa=\sqrt{-q_x^2}$ is a real positive parameter. In this situation, the propagating solutions given in (\ref{21}-\ref{22}) are replaced by evanescent modes. The corresponding wave functions therefore take exponential forms proportional to $e^{\pm \kappa x}$, describing spatially decaying or increasing solutions. Consequently, the transmission and reflection coefficients can be expressed in terms of hyperbolic functions instead of trigonometric ones
		For the case  $q_x^2 < 0$ (evanescent modes), These coefficients become
		\begin{align}
			&t=\frac{e^{-i  D k_x }
				\cos\phi\cos\theta }{
				\cos\phi\cos\theta\cosh(D \kappa ) -  \sinh(D \kappa ) (1-\sin\phi\sin\theta)
			}\\
			&r=\frac{ie^{i\phi}\sinh( D\kappa )(\sin\theta-\sin\phi)}{
				\cos\phi\cos\theta\cosh( D \kappa ) - \sinh( D \kappa ) (1-\sin\phi\sin\theta)}.
		\end{align}
		These evanescent solutions play an important role in describing tunneling processes and ensure the continuity of the wave function and its derivatives at the interfaces.

	To determine the transmission and reflection probabilities, we use the continuity equation for the current density to derive explicit expressions for the currents associated with the different wave components. Specifically, the incident, reflected, and transmitted current densities are given by
	\begin{align}
		&	J_{i}=2v_Fcos(\phi) \\
		&J_{r}=2v_Fr^*rcos(\phi)\\
		&J_{t}=2v_Ft^*tcos(\phi).
	\end{align}
	These quantities provide a rigorous framework for evaluating the   probability flow across the system. By taking the ratio of the transmitted to the incident current densities,  $T=\frac{|J_{t}|}{|J_{i}|}=|t|^2$, 
	we obtain the transmission probability
\begin{align}
			T=\frac{
				\cos^2\phi\cos^2\theta}{
				\cos^2\phi\cos^2\theta\cos^2(D q_x ) +\sin^2(D q_x ) (1-\sin\phi\sin\theta)^2
			}
	\end{align}
	which directly connects the transmission amplitude to the measurable probability of wave propagation through the barrier. A similar procedure yields the reflection probability, ensuring probability conservation within the system ($R=1-T$).
	
	At zero temperature, the conductance is defined as the average flux of fermions across half of the Fermi surface \cite{butker, conduct1}, emphasizing that only the occupied states at the Fermi level contribute to transport. Alternatively, the conductance can be expressed in terms of the transmission properties of the system. In this formulation, it is given by the integral of the total transmission over the transverse momentum 
	$k_y$ \cite{Biswas}, highlighting the contribution of all possible transverse channels to the net conductance. Mathematically, the conductance is obtained by summing the contributions of each channel, weighted by its transmission probability. This yields
	\begin{align}
		G_\tau &= G_0 \int_{-k_y^{\text{max}}}^{k_y^{\text{max}}} (T_{\uparrow\tau}(E, k_y)+T_{\downarrow\tau}(E, k_y))\, dk_y	\\
		&=G_0 g_{\tau}
	\end{align}
	where $G_0$ denotes the conductance unit, $E_F$ is the Fermi energy,  and $k_y^{\text{max}}$ represents the maximum wave vector component along the $y$-direction. Taking into account that $k_y^{\text{max}} = k \sin\phi^{\text{max}}$, the conductance  can be expressed as 
	\begin{align}
		G_\tau = G_0 \int_{-\phi^{\text{max}}}^{\phi^{\text{max}}} (T_{\uparrow\tau}(E, \phi)+T_{\downarrow\tau}(E, \phi)) \cos\phi \, d\phi
	\end{align} 
	and $\phi^{\text{max}}$ is the maximum incident angle. 
	This relation connects the microscopic transmission amplitudes, obtained from the wave-function matching procedure, to the macroscopic conductance, making it possible to relate the quantum transport properties of the system to experimentally measurable observables.

	Spin and valley polarization provide quantitative measures of the imbalance in electronic transport between different spin orientations and valley degrees of freedom. Since electrical conductance depends on both the availability of propagating states and their transmission probabilities through the structure, polarization effects naturally emerge when these quantities become spin- or valley-dependent. Such asymmetries can be induced by external fields, intrinsic spin-orbit coupling, or symmetry-breaking potentials, and they play a central role in spintronic and valleytronic applications \cite{polar}.
	In this framework, spin polarization is defined by comparing the conductance contributions from the spin-up and spin-down channels, whereas  
	valley polarization is obtained by contrasting the conductance associated with the two inequivalent valleys, $K$, $K'$. 
	These quantities are expressed as normalized differences, ensuring that polarization remains between -1 and 1. Total conductance serves as the normalization factor. They are
	\begin{align}
		&P_s=\frac{G_\tau(\uparrow)-G_\tau(\downarrow)}{G_\tau(\uparrow)+G_\tau(\downarrow)}\\
		&P_v=\frac{G_\tau-G_{-\tau}}{G_\tau+G_{-\tau}}.
	\end{align}
	Note that nonzero values of $P_s$ or $P_v$ indicate preferential transport through specific spin or valley channels, reflecting the underlying symmetry breaking in the system. In particular, strong magnetic fields can lift spin and valley degeneracies, thereby enabling selective control of fermionic transport.

	\section{NUMERICAL RESULT}\label{Num}

	We present and discus our numerical results in detail. We analyze the transmission and reflection coefficients, the conductance, and the spin and valley polarizations in order to clarify how the system parameters affect these key physical quantities and to uncover the mechanisms governing electronic transport. The energy spectrum is first examined, as it provides valuable insight into the available electronic states and their evolution under applied fields. We then focus on transmission and conductance to assess the efficiency of carrier transport across the barriers. Finally, the analysis of spin and valley polarizations illustrates how these internal degrees of freedom can be selectively manipulated. Together, these results provide a comprehensive picture of confinement effects and transport modulation in two-dimensional materials. They also point toward promising applications in spintronics and valleytronics.

	\begin{figure}[ht!]
		\centering
		\subfloat[]{\centering\includegraphics[scale=0.55]{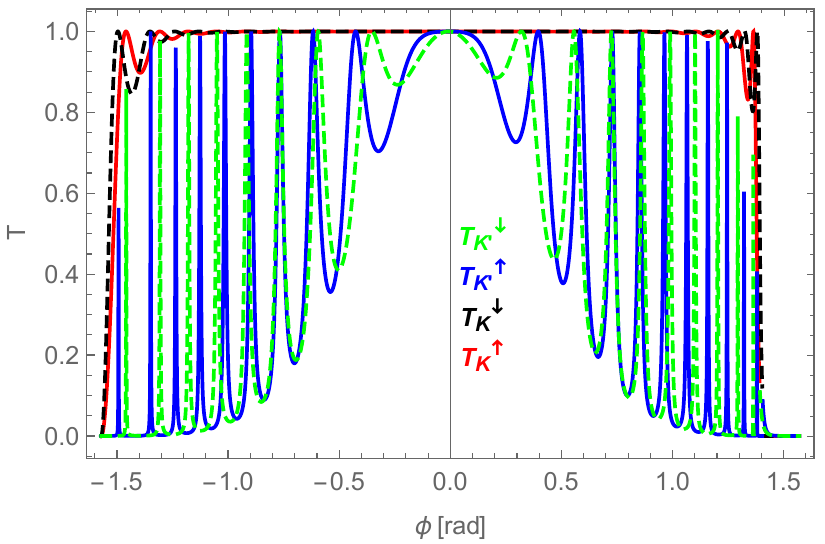}\label{fig2a}}\\
		\subfloat[]{\centering\includegraphics[scale=0.55]{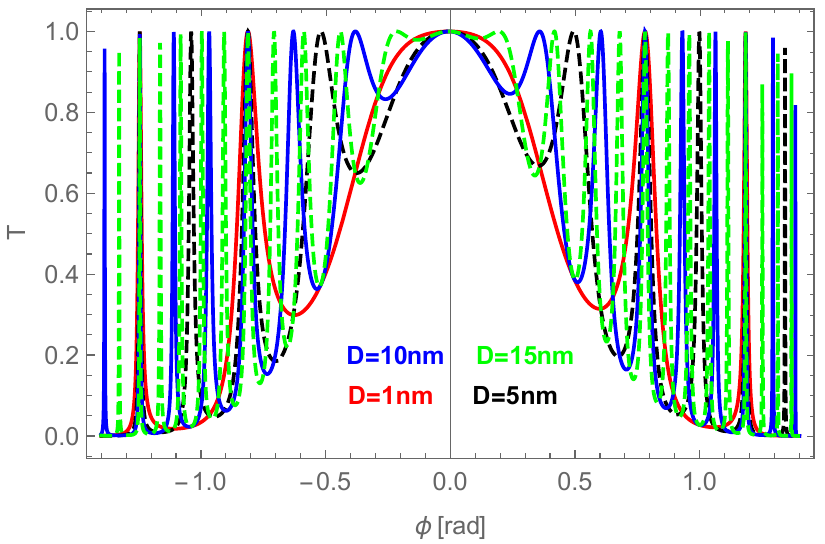}\label{fig2b}}\\
		\subfloat[]{\centering\includegraphics[scale=0.55]{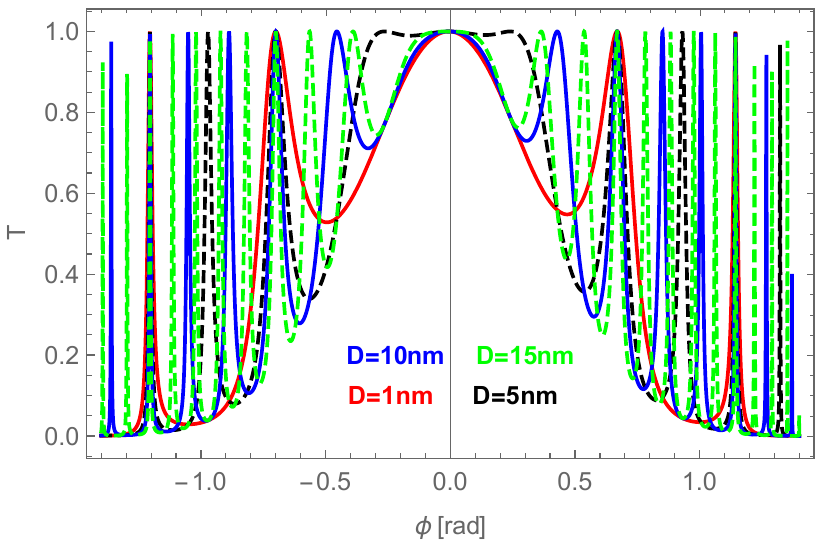}\label{fig2c}}\\
		\caption{Transmissions as a function of incident angle for $B=6$ T and $E=0.95$ eV. 
			(a): Spin-up and spin-down transmission for the $K$ and $K'$ valleys with $D = 5$ nm, (b): Spin-up transmission for the $K$ valley, (c): Spin-down transmission for the $K'$ valley.} \label{fig2}

	\end{figure}
	
	Figure~\ref{fig2} shows the transmission as a function of the incidence angle for a magnetic field of intensity $B = 6$~T. 
	The transmission exhibits an oscillatory behavior, displaying a clear Klein tunneling effect \cite{klien1,klien2}, {as it remains perfect at normal incidence in all cases.} Fig.~\ref{fig2a} illustrates the transmission behavior for both valleys in spin-up and spin-down states. For valley $K$, the transmission is nearly perfect for all incidence angles between $-\pi/2$ and $\pi/2$, with oscillations appearing near the boundaries of this interval
	For valley $K^\prime$, the transmission oscillates between 0 and 1
	but remains perfect at small angles. The spin-up and spin-down transmissions in this valley are almost identical, with a slight relative shift. Fig.~\ref{fig2b} presents the spin-up transmission for valley $K^\prime$. 
	Increasing the barrier width modifies the transmission and leads to additional peaks of perfect transmission. 
	The same behavior is observed in Fig.~\ref{fig2c}. Thus, increasing the barrier width increases the number of peaks,  as it provides fermions with more opportunities to interact with the magnetic field. 
	This interaction modifies the fermion energy, resulting in the emergence of bound and quasi-bound states. The interference between these states give rise to Fano resonances \cite{Fano}, which modulate the transmission probability and enhance confinement. 
	{In summary,  the barrier width and the incident angle are identified as the key parameters for guiding fermions and selecting which transmission channel crosses the barrier. Since the magnetic field remains constant in this case, its effect cannot be directly inferred from these data and is therefore not included in this discussion.}

	\begin{figure}[ht]
		\centering
		\subfloat[]{\centering\includegraphics[scale=0.55]{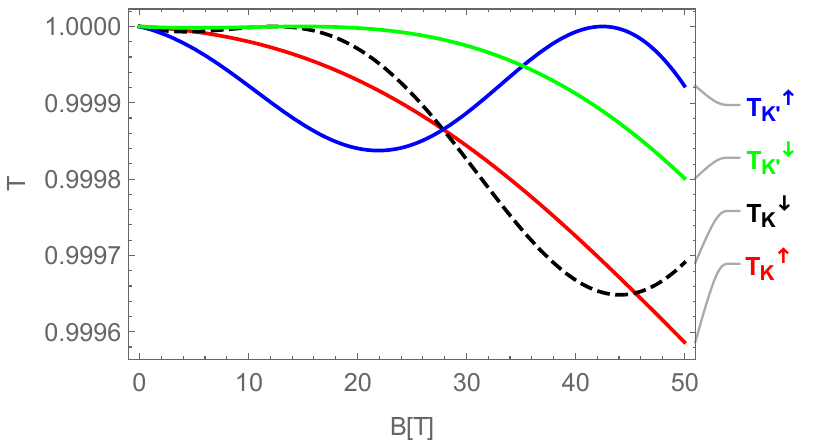}\label{fig1a}}\\
		\subfloat[]{\centering\includegraphics[scale=0.55]{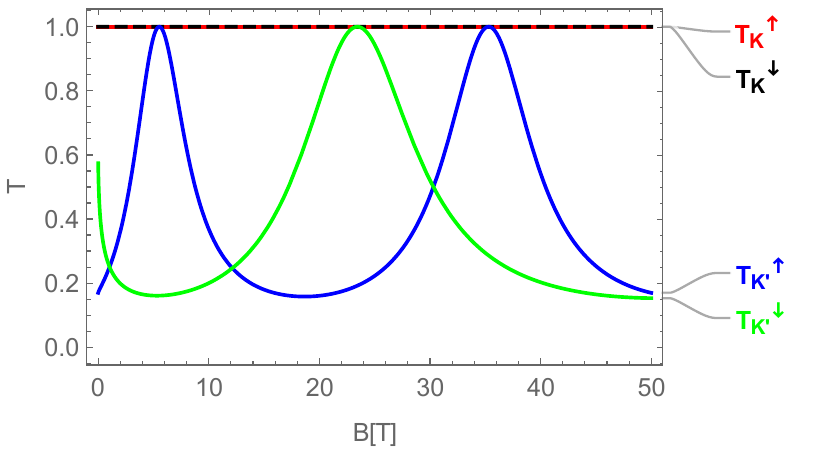}\label{fig1b}}\\
		\subfloat[]{\centering\includegraphics[scale=0.55]{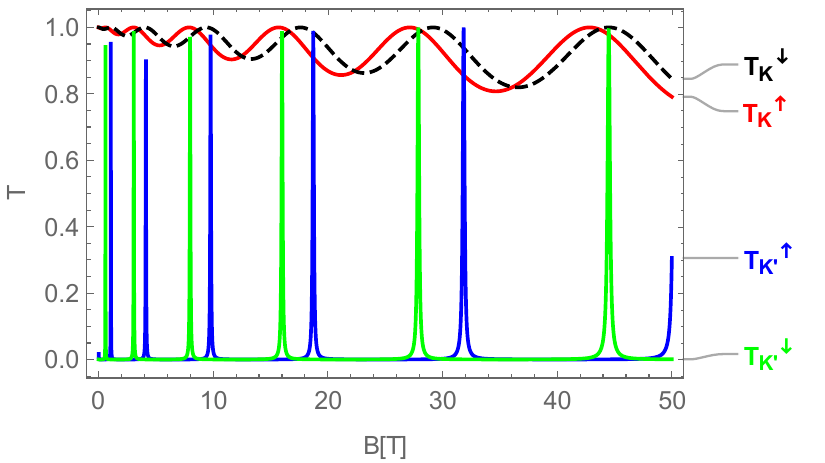}\label{fig1c}}
		\caption{
			{Spin-up and spin-down transmission as a function of the magnetic field $B$ for the $K$ and $K'$ valleys with $E = 1.2$ eV,  $D=15$ nm, and  three incident angles} (a): $\phi=0$, (b): $\phi=15^\circ$,  (c): $\phi=30^\circ$.
		}\label{fig1} 
	\end{figure}

	Figure~\ref{fig1} summarizes the evolution of the transmission probability as a function of the magnetic field $B$ for different incidence angles and valley indices. At normal incidence, $(\phi = 0)$, shown in Fig.~\ref{fig1a}, the transmission curves remain essentially flat as the magnetic field increases. This behavior indicates that electrons incident normally cross the barrier with nearly perfect efficiency, and that the magnetic field has a negligible effect on transport in this regime.
	When the incidence angle is increased to $(\phi = 15^\circ)$, as illustrated in Fig.~\ref{fig1b}, qualitative changes begin to appear. In particular, electrons in the $K'$ valley become noticeably more sensitive to the magnetic field and exhibit clear oscillations in the transmission. In contrast, the transmission associated with the $K$ valley remains largely unaffected over the same range of $B$. This marked difference between the two valleys indicates
	the emergence of  magnetic-field-induced valley selectivity, which becomes active once the condition of normal incidence is relaxed. 
		The pronounced transmission peak observed around $\theta \approx 15^\circ$ for both $K$ and $K'$ valleys can be attributed to a resonant transmission condition. At this incidence angle, the longitudinal and transverse components of the wave vector satisfy the matching conditions across the interfaces, leading to constructive interference of the propagating modes inside the barrier. As a result, the tunneling probability is strongly enhanced and the transmission approaches unity. This behavior is analogous to resonance-assisted transport mechanisms reported in the literature, where specific incidence angles allow nearly perfect transmission through potential barriers due to phase matching of the electronic wave functions. A similar phenomenon has been discussed in \cite{Coulomb}.
	At a larger angle, $(\phi = 30^\circ)$, shown in Fig.~\ref{fig1c}, the influence of the magnetic field is further enhanced. In the $K'$ valley, transmission is strongly suppressed over most magnetic field values, except at isolated points where perfect transmission is recovered, indicating resonant tunneling through the barrier.  
	By comparison, electrons in the $K$ valley continue to transmit more efficiently overall. Although their transmission decreases with increasing $B$ and develops an oscillatory structure, it remains consistently higher than that of the $K'$ valley.
	Taken together, these results confirm that the magnetic field primarily affects electron transport at oblique incidence, while transmission at normal incidence remains remarkably robust. Such behavior is characteristic of two-dimensional materials, including graphene and molybdenum disulfide, where normal-incidence transmission is known to be protected against external perturbations. Our findings are fully consistent with this general picture in good agreement with our previous studies \cite{Elaitouni2022,Elaitouni2023A,laser+mag,doublelaser+magn}.
		Here, we note that physically, the isolated transmission resonances where $T=1$ originate from {quantum interference and resonant tunneling} in the $K'$ channel. In such systems, an applied magnetic field quantizes the charge carriers' motion into discrete cyclotron orbits (Landau levels), establishing standing-wave conditions inside the barrier or scattering region. When the accumulated phase of the electron wavefunction satisfies the constructive interference condition, effectively aligning the energy of electron with a discrete resonant state, perfect transmission occurs, resulting in $T=1$ peaks reminiscent of Fabry--P\'erot-type resonances observed in graphene-based structures under magnetic field modulation~\cite{Tworzydlo2006}.  
		Furthermore, as the magnetic field increases, the {spacing between resonances also increases} because the effective phase accumulated per unit change in magnetic field becomes larger due to the increasing separation of the Landau-quantized states. In graphene and other Dirac materials, Landau levels are not equally spaced and their energies scale nonlinearly with magnetic field, so higher fields cause larger shifts in the resonance condition for successive modes~\cite{prop}.

	
	\begin{figure}[ht!]
		\centering
		\subfloat[]{\centering\includegraphics[scale=0.55]{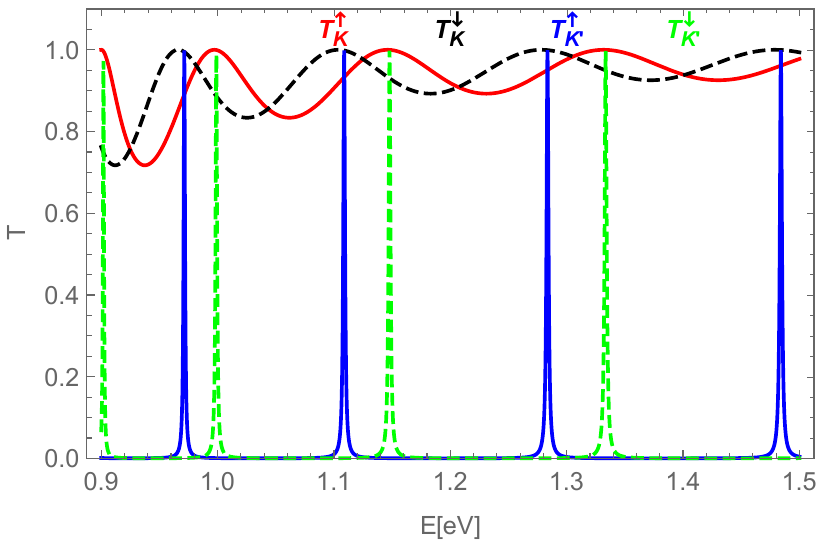}\label{fig0a}}\\
		\subfloat[]{\centering\includegraphics[scale=0.55]{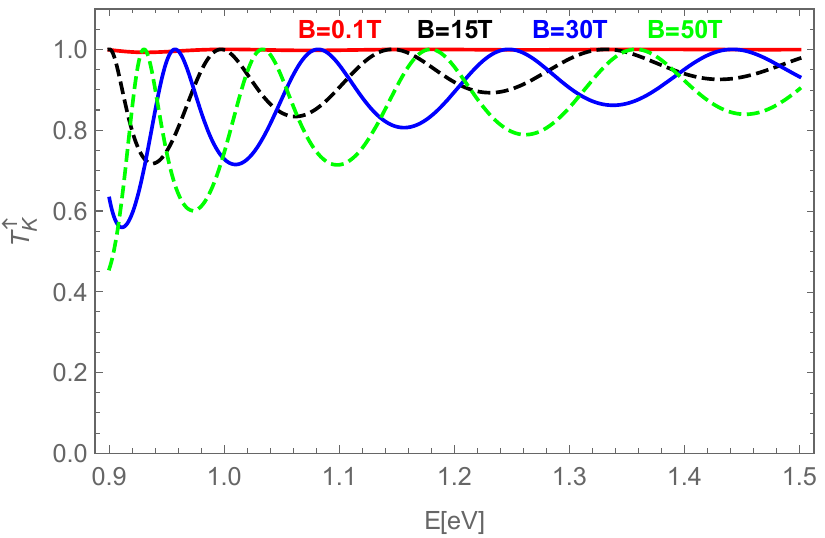}\label{fig0b}}\\
		\subfloat[]{\centering\includegraphics[scale=0.55]{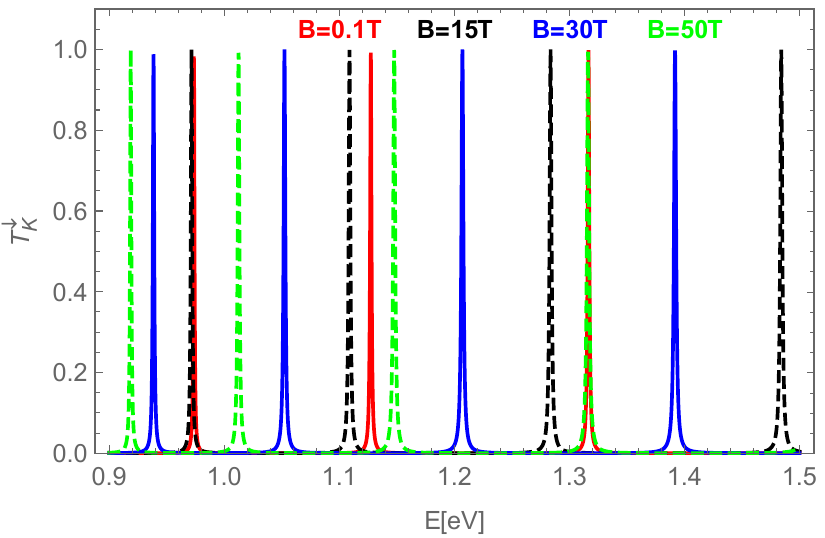}\label{fig0c}}
	\caption{Transmission as a function of the incident energy $E$ for oblique incidence  $\phi = 30^\circ$ with $D=5$ nm. 
				(a): Spin-up and spin-down transmission for the $K$ and $K'$ valleys with $B = 15$ T, 
				(b): Spin-up transmission, and 
				(c) Spin-down transmission for the $K$ valley and four values of $B$. }\label{fig0} 
	\end{figure}
	
	Figure~\ref{fig0} shows  the transmission of spin-up and spin-down as function of the incident energy $E$
	for  the $K$ and $K'$ valleys. In Fig.~\ref{fig0a}, the transmission  for both spin orientations and both valleys are plotted together. A clear asymmetry between the two valleys is immediately apparent: electrons in the $K$ valley transmit much more efficiently than those in the $K'$ valley. Over a broad range of incident energies, transmission in the $K'$ valley is almost completely suppressed, while the two spin components display a small phase shift with respect to each other. This behavior suggests that the magnetic field $B$ effectively modifies the valley orientation, favoring transport through one valley channel over the other.
	The effect of $B$ becomes more evident when focusing on the spin-up transmission in the $K$ valley, as shown in Fig.~\ref{fig0b}. For a weak magnetic field (red line), the transmission remains close to unity. This means that fermions move through the barrier  almost as if it were absent. However, as the $B$ increases, striking changes occur. The transmission no longer remains flat, instead, oscillations emerge and become more pronounced as $B$ increases, while the total transmission slowly decreases. This signals a richer, more complex interplay between the magnetic field and the quantum tunneling process.
	Fig.~\ref{fig0c} focuses on the spin-up transmission in the $K'$ valley. Here, transmission is nearly turned off across most incident energies and only turns on at a few sharp, isolated points where resonant tunneling occurs. Increasing $B$ 
	{dramatically modifies the tunneling spectrum, making transmission highly sensitive to both the energy of the incoming particles and the strength of the magnetic field.}
	This indicates that the interaction between the field and the fermions alters their effective orientation and enhances valley polarization. Consequently, for sufficiently strong fields, most fermions originating from the $K$ valley can cross the barrier, while only a small fraction of fermions from the $K'$ valley can transmit.

	\begin{figure}[ht]
		\centering
		\subfloat[]{\centering\includegraphics[scale=0.52]{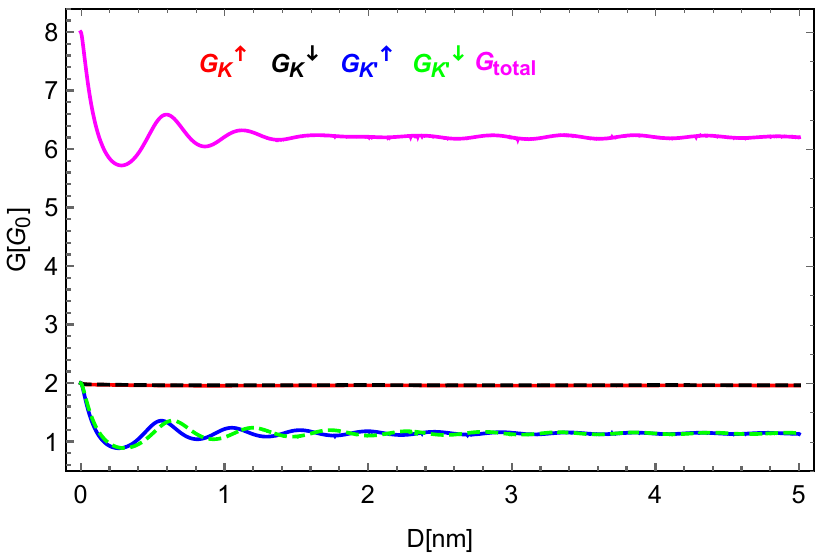}\label{fig4a}}\\
		\subfloat[]{\centering\includegraphics[scale=0.52]{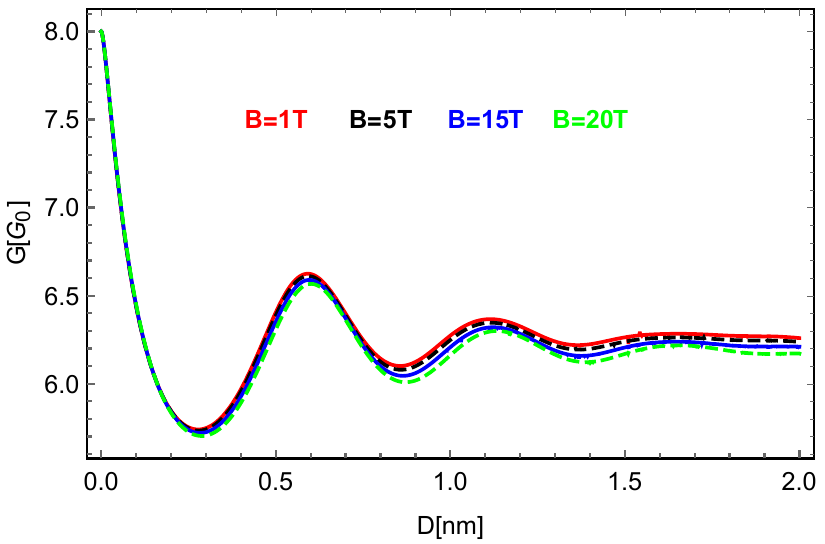}\label{fig4b}}\\
		\subfloat[]{\centering\includegraphics[scale=0.52]{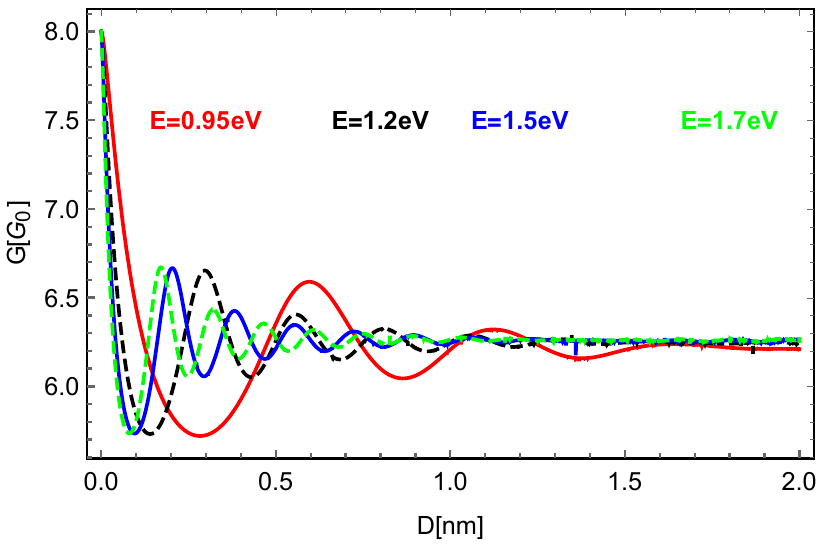}\label{fig4c}}
		\caption{Conductance as a function of the barrier width $D$. (a):  $E=0.95$ eV and $B=15$ T, (b):  $E=0.95$ eV and four values of $B$, (c):  $B=15$ T and  four values of $E$.}\label{fig4}
	\end{figure}
	

	Figure~\ref{fig4} shows the conductance as a function of barrier width $D$ for the four transmission channels. Fig.~\ref{fig4a} indicates that the conductance in the $K$ valley  remains nearly constant at the  value 2, consistent with previous results (Figs.~\ref{fig2}-\ref{fig1}), since transmission in  $K$ is more likely than in  $K'$. In contrast, for the $K'$ valley, the conductance decreases in an oscillatory manner before stabilizing at its minimum value. The total conductance reaches its maximum at the smallest barrier width, reflecting the nearly perfect transmission for very thin barriers.
	Fig.~\ref{fig4b} presents the total conductance for four different values of the applied magnetic field. It is observed that the conductance decreases as the magnetic field increases. This behavior arises because the magnetic field modifies  the energy levels of the fermions, confining them between discrete levels created by the field, a phenomenon related to the Zeeman effect~\cite{Zemman}.
	Fig.~\ref{fig4c} shows the conductance for four values of the incident energy $E$. 
	Increasing $E$ reduces the amplitude of the oscillations and stabilizes the conductance, as higher energies allow the fermions to cross the barrier more easily. It is clearly seen that increasing the barrier width $D$ decreases the conductance until it stabilizes at an average value, after which further variations in 
	$D$ have little effect. Similarly, increasing the magnetic field reduces the conductance due to its interaction the fermions, while higher incident energies accelerate the stabilization of the conductance.

	\begin{figure}[ht]
		\centering
		{\centering\includegraphics[scale=0.55]{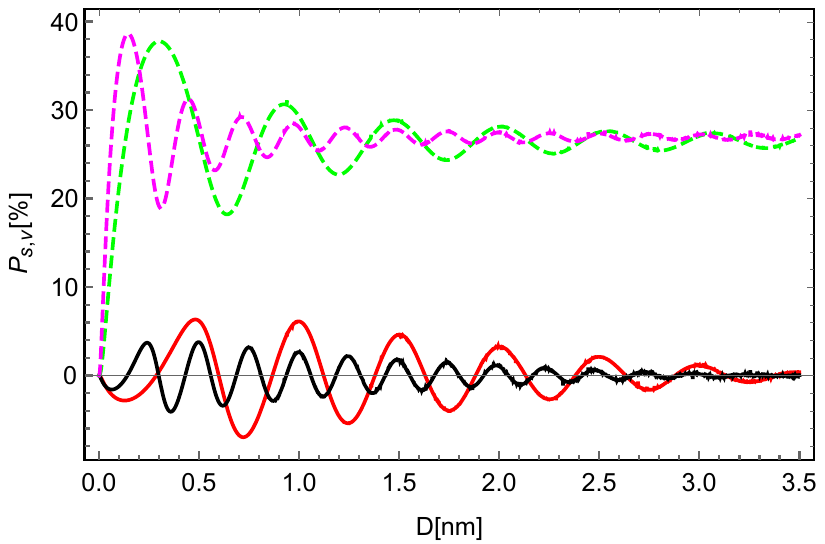}\label{fig5a}}
		\caption{Polarization as a function of barrier width $D$ for $B = 15$ T. Spin polarization ($E=0.92$ eV: red line, $E=1.2$~eV: black line), and valley polarization ($E=0.92$~eV: magenta dashed line, $E=1.2$ eV: green dashed line).}\label{fig5} 
	\end{figure}
	
	In Figure~\ref{fig5}, we show two values of incident energy and the resulting spin and valley polarization (solid and dashed lines, respectively). At a magnetic field of ($B = 15$ T), spin polarization remains close to zero, indicating that there is no favoring of either spin direction during the transmission. Upon increasing incident energy, amplitude of spin oscillations decreases, meaning higher energy fermions are less influenced by spin (or spin selection), and consequently, spin effects become   negligible. 
	In contrast, valley polarization exhibits a different behavior. We see positive oscillations around a mean of~ $25\%$, demonstrating a clear preference to one valley when fermions cross the barrier. Unlike spin polarization, valley polarization is less sensitive to higher incident energies, highlighting the selective nature of the system response. 
	This selective valley transport indicates that the system can function as a valley filter, representing a first step toward controlled manipulation of the valley degree of freedom.
	These findings are particularly significant for valleytronics, as they demonstrate that valley polarization can be generated and maintained through barrier engineering and external fields. The ability to achieve sizable and stable valley polarization while suppressing spin polarization opens promising opportunities for devices that exploit valley degrees of freedom for information processing and filtering applications.


	\section{Conclusion}\label{Concl}
	
	We analyzed how a magnetic field influences the motion of fermions in tungsten diselenide (WSe$_2$), a two-dimensional material with particularly interesting spin and valley properties. By introducing a magnetic barrier with a finite width \(D\), the system is divided into three regions, each experiencing a different magnetic environment. To understand the transport process, we solve the eigenvalue equation in each region and obtain the corresponding eigenspinors for the two valleys. By matching these eigenspinors at the boundaries, we calculate the transmission and reflection probabilities. From the resulting current densities, we determine the total transmission, which allows us to evaluate the conductance at zero temperature. This procedure also makes it possible to extract both spin and valley polarizations, leading to  direct link the microscopic behavior of the fermions to observable transport properties.
	The magnetic field plays a key role in shaping the fermion dynamics. It modifies the energy levels of the carriers and can partially trap them  inside the barrier. At the same time, the field bends their trajectories and introduces phase differences in their eigenspinors. These phase effects can interfere with one another, often  destructively,  which reduces the probability that fermions will successfully cross the barrier. As a result, transport through the system becomes strongly dependent on the  magnetic field and  the barrier width.

	One of the most important findings  is the strong influence of the magnetic barrier on the valley degree of freedom. While spin polarization remains relatively weak in within the considered parameter range, valley polarization exhibits a clear and robust response.
	We find that valley polarization can exceed $30\%$, meaning that near fermions transmitted through the barrier come from the comparable valley. 
	In practical terms, the magnetic barrier selectively favors one valley over the other, effectively functioning as a valley filter that guides fermions along a desirable path.
	An additional advantage of this behavior is its tunability.
	By adjusting the attractive field strength, the barrier width, or the incident energy, the degree of valley polarization can be controlled over a considerable range. 
	This level of control is critical for practical applications, as it allows the transport properties to be tailored without modifying the material itself. Furthermore, the fact that valley polarization remains stable under different conditions suggests that this effect should be realizable in actual devices.
	
	
	Our results highlight the potential of attractive magnetic barriers as effective tools for manipulating valley transport in two-dimensional materials. The ability to generate and control valley polarization opens promising opportunities for valleytronics, where information can be encoded and processed using valley states instead of charge alone. By demonstrating how attractive fields can be used to steer fermions in WSe$_2$, this work provides valuable insights for future research and lays the foundation for the development of valley-based electronic and memory devices.

\end{document}